\RequirePackage{snapshot}
\documentclass[3p,twocolumn,number,sort&compress,12pt]{elsarticle}

\usepackage{cmap}
\usepackage{amssymb}
\usepackage{lineno}
\usepackage[multidot]{grffile}
\usepackage{textcomp}
\usepackage[utf8x]{inputenc}
\usepackage{url}
\usepackage{xcolor}

\graphicspath{{./figures/}}
\newcommand{\GH}[1]{#1}

\begin{document}
   \sloppy
   \begin{frontmatter}

      \title{Visualization of steps and surface reconstructions in
	 Helium Ion Microscopy with atomic precision}

      \author[UT]{Gregor Hlawacek\corref{cor}\fnref{present,HIM}}
      \ead{g.hlawacek@hzdr.de}
      \cortext[cor]{Corresponding author}
      \fntext[present]{Present address: Ion beam Physics and Materials Research,
	 Helmholtz---Zentrum Dresden -- Rossendorf, Bautzner Landstraße 400, 01328,
	 Dresden, Germany}
      \fntext[HIM]{Acquired and analyzed all HIM based data. AFM analysis. Wrote the manuscript.}

      \author[UT]{Maciej Jankowski\fnref{Grenoble,LEEM}}
      \fntext[Grenoble]{Present address: ID03 Surface Diffraction Beamline,
	 ESRF--The European Synchrotron, 71 Avenue des Martyrs, 38400 Grenoble, France}
      \fntext[LEEM]{Acquired and analyzed all LEEM based data.}

      \author[UT]{Herbert Wormeester\fnref{supervisor}}
      \author[UT]{Raoul van Gastel\fnref{supervisor}}
      \author[UT]{Harold J. W. Zandvliet\fnref{supervisor}}
      \author[UT]{Bene Poelsema\fnref{supervisor}}
      \fntext[supervisor]{Helped in interpreting the results and ensured financial support.}

      \address[UT]{Physics of Interfaces and Nanomaterials, University of Twente,
	 PO Box 217, 7500 AE, Enschede, The Netherlands}

      \begin{abstract}
	 Helium Ion Microscopy is known for its surface sensitivity and high
	 lateral resolution. Here, we present results of a Helium Ion
	 Microscopy based investigation of a surface confined alloy of Ag on
	 Pt(111). Based on a change of the work function of 25\,meV across
	 the atomically flat terraces we can distinguish Pt rich from Pt
	 poor areas and visualize the single atomic layer high steps between
	 the terraces. Furthermore, dechanneling contrast has been utilized
	 to measure the periodicity of the hcp/fcc pattern formed in the
	 2--3 layers thick Ag/Pt alloy film. A periodicity of 6.65\,nm along
	 the $\langle\overline{11}2\rangle$ surface direction has been
	 measured. In terms of crystallography a hcp domain is obtained
	 through a lateral displacement of a part of the outermost layer by
	 $1/\sqrt{3}$ of a nearest neighbour spacing along
	 $\langle\overline{11}2\rangle$. This periodicity is measured with
	 atomic precision: coincidence between the Ag and the Pt lattices is
	 observed for 23 Ag atoms on 24 Pt atoms. The findings are perfectly
	 in line with results obtained with Low Energy Electron Microscopy
	 and Phase Contrast Atomic Force Microscopy.
      \end{abstract}


   \end{frontmatter}


   \section{Introduction}

   Helium Ion Microscopy (HIM)~\cite{Hlawacek2013c,Joy2013} has become a
   powerful imaging tool with very high lateral resolution and surface
   sensitivity. With a lateral resolution better than
   0.5\,nm~\GH{\cite{Hill2011}} one would hope
   that also information on the \GH{lattice} structure of a sample surface can be
   obtained using this method. So far this has not been demonstrated.
   However, results on several atoms high steps and surface termination of
   Ti$_3$SiC$_2$~\cite{Buchholt2011} show that this goal is not out of
   reach. In addition, theoretical reports indicate that imaging of the
   atomic structure should in principle be possible for very thin
   layers~\cite{Zhang2012}. In a real experiment one would have to care
   about vacuum levels to minimize hydrocarbon contamination and---in
   particular on thick samples---damage by the \GH{recoiling
      substrate atoms} as
   well as the implanted Helium. So far this has prevented the observation
   of features related to the atomic structure of the sample surface.
   
   Here, we present\GH{---the to the best of our knowledge---}first observation of single
   atom high steps with a Helium Ion Microscope. In addition we will demonstrate
   that under specific conditions the HIM is able to distinguish areas in which
   a small number of atoms have been moved from the bulk lattice position by
   a fraction of an inter atomic spacing. 

   Since its introduction by Ward, Notte, and Economou~\cite{Economou2006}
   Helium Ion Microscopy (HIM) has become an important microscopy technique
   providing high resolution images of sample surfaces. This is true for
   conducting as well as insulating materials. The present work is based on
   the well known image formation mechanisms in HIM which utilize secondary
   electrons excited by the primary ion and ejected from the sample
   surface~\cite{Petrov2011,Cazaux2010a,Lin2005}. It relies on the high
   surface sensitivity~\cite{Hlawacek2013b} of the tool and the fact that
   channeling can be exploited to enhance the imaging of thin surface
   layers~\cite{Hlawacek2012}.

   Low Energy Electron Microscopy (LEEM)~\cite{Bauer1998} and Atomic Force
   Microscopy (AFM)---in particular phase contrast
   AFM~\cite{Whangbo1998,Schmitz1997,Garcia2007}---have been used to
   benchmark our findings. 
   
   The Ag/Pt(111) system is a representative example of a surface confined
   alloy, which is widely studied in the field of surface
   science~\cite{Becker1993,Roeder1993,Strueber1993,Zeppenfeld1995,Tersoff1995,Jankowski2014a}.
   Deposition under UHV conditions of a 2--3 layers\footnote{Coverage is
      given in mono layer equivalents based on the Pt(111) surface unit
      cell.} thick Ag film on Pt(111) at room temperature followed by an
   annealing step above 550\,K results in irreversible changes of the
   surface morphology~\cite{Becker1993}. These changes were identified by
   Brune et al.~\cite{Brune1994} using Scanning Tunneling Microscopy (STM)
   as the formation of a well-ordered periodic dislocation network, formed
   through material intermixing in the first two layers of the deposited
   film~\cite{Bendounan2012,Ait-Mansour2012}. The symmetry, periodicity and
   level of ordering of the network depends on the magnitude of material
   intermixing~\cite{Jankowski2014} which is controlled mainly by the
   substrate temperature during deposition. The structural model of the
   network formed at 800\,K was revised recently by A\"it--Mansour et
   al.~\cite{Ait-Mansour2012}. According to their model, the deposition of a
   2 layers thick Ag film on Pt(111) followed by annealing at 800\,K leads
   to the exchange of atoms between Pt(111) and the deposited Ag layer. The
   Pt interface layer contains Ag inclusions, whereas the expelled Pt atoms
   from the substrate form inclusions in the top layer of the alloy. The
   mixing of the atoms is limited only to fcc--stacking sites. Stacking
   faults are present at the Pt interface layer and the formed network has a
   3--fold symmetry. Further deposition of Ag leads to the growth of a third
   layer which was reported to be purely silver and hexagonal in
   structure~\cite{Ait-Mansour2012}. The formed dislocation network at the
   alloy--Pt(111) interface causes periodical undulations of the third and
   subsequent Ag layers~\cite{Jankowski2014}.

   \section{Experimental}

   Helium Ion Microscopy has been performed in an ultra high vacuum (UHV)
   HIM~\cite{Veligura2012,vanGastel2011} Orion$^+$ from Carl Zeiss
   Microscopy. The system has a base pressure of $2\times10^{-9}$\,mbar.
   This pressure is reached thanks to a stainless steel sample chamber with
   Conflat type flanges, a modified pumping and load lock strategy, a
   5000\,l/s titanium sublimation pump and a differentially pumped door
   gasket. The system is equipped with a standard Everhardt Thornley
   Detector to record secondary electron (SE) based images. In addition,
   detectors to count back scattered helium
   (BSHe)~\cite{Hlawacek2012,Gastel2012} measure their
   energy~\cite{Sijbrandij2010,Behan2010} and collect photons to enable
   ionoluminescence~\cite{Veligura2014b,IL_Boden} studies of materials are
   present. The presented images were recorded using a sample tilt of
   35\textdegree{} to exploit channeling into the underlying bulk crystal.
   This is necessary to maximise the surface contrast~\cite{Hlawacek2013b}. 

   The samples were prepared and initially characterized in a Low
   Energy Electron Microscope (LEEM) Elmitec III. The used Pt(111) crystal
   had a miscut angle of less than 0.1\textdegree. Surface cleaning was
   done by prolonged repetitive cycles of argon ion bombardment,
   annealing in 2x10$^{-7}$\,mbar of oxygen at 800\,K, and subsequent
   flashing to 1300\,K. 

   High purity silver (99.995\%) was deposited from a molybdenum
   crucible mounted in an electron beam evaporator (Omicron EFM-3). The
   growth of the silver layers was tracked in-situ and real time using
   bright--field mode~\cite{Tromp1993,Bauer1994} in LEEM.

   Venting and pumping of the vacuum systems has been timed to minimize
   contamination of the sample surface while transporting it from the LEEM
   to the HIM vacuum chamber. However, adsorption of hydro carbons on the
   film surface is most likely and could not be prevented. Since oxidation
   is strongly suppressed by the presence of silver~\cite{Jankowski2014a}
   oxygen is considered a lesser problem. The sample has been imaged in the
   LEEM after the HIM analysis has been completed. Comparing the results to
   data recorded before the sample transport showed no significant changes
   relevant for the present study.
   
   The AFM measurements were done under ambient conditions with an Agilent
   5100 AFM employing amplitude--modulation to record the topography. A
   MikroMasch Al back--coated NSC35 Si$_{3}$N$_{4}$ cantilever with a tip
   radius of 8\,nm was used in these measurements. The resonance frequency
   of this cantilever type is 205\,kHz and the nominal spring constant is
   8.9\,N/m. For the measurements an amplitude set-point of 90\% was used
   and the oscillation amplitude was in the range between 30\,nm and 40\,nm.

   \section{Results}

   \subsection{Surface mounds and HIM sample alignment} 

\begin{figure} \centering
      \includegraphics[width=\linewidth]{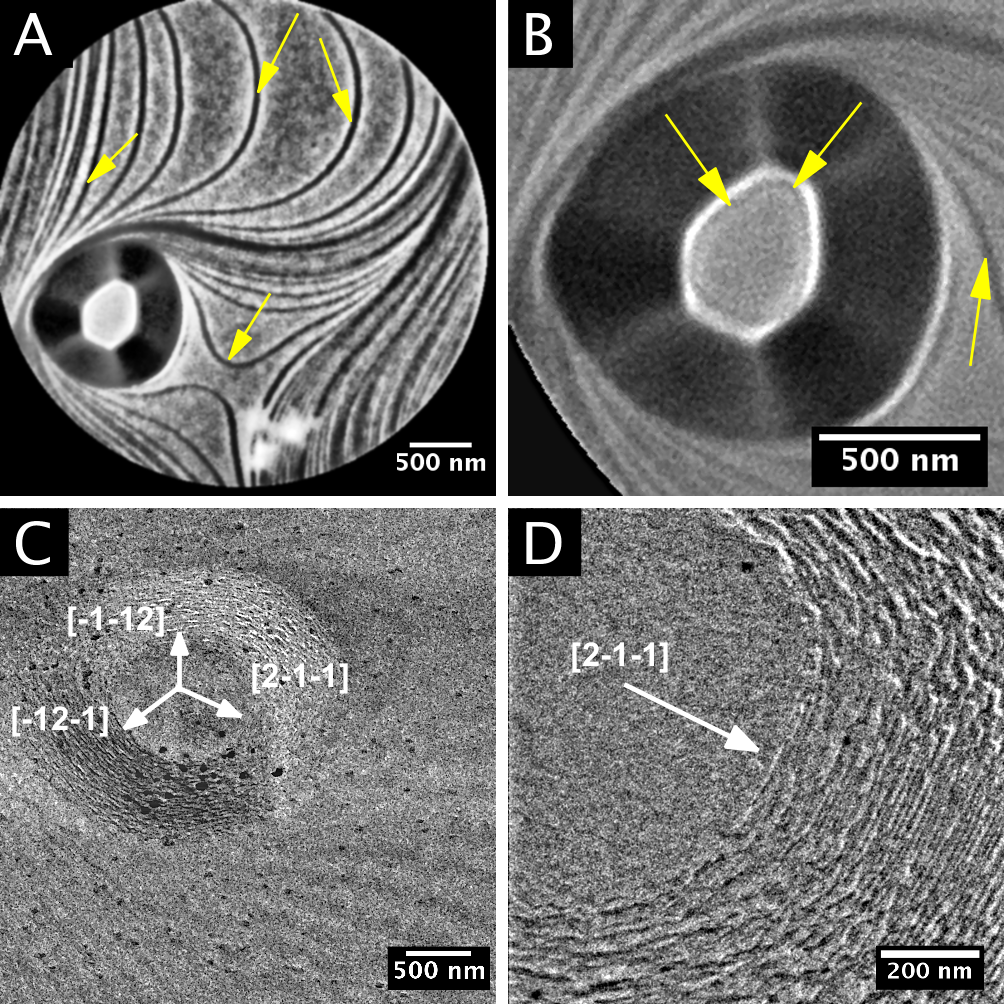}
      \caption{\label{fig:LEEM-HIM} (A) Representative LEEM image of a mound
	 on the clean Pt(111) surface. The curved lines (a few marked by
	 yellow arrows) are monatomic steps on the sample surface (electron
	 energy 2\,eV). The three white spots at the bottom of the image are
	 damage on the micro channel plate used to record the images. (B)
	 Magnification of the mound. The preponderingly hexagonal shape of
	 the mound is clearly visible. A monatomic surface step on the top
	 of the mound is marked with yellow arrows. (C) HIM image of a Pt
	 mound on single crystalline Pt(111) sample covered by 2--3 layers
	 equivalent of Ag. White arrows indicate the threefold symmetry of
	 the steps with $\left\{001\right\}$ facet orientations. The labels
	 give the projected surface directions in the (111) plane. (D) High
	 resolution HIM image of a mound.} 
   \end{figure}

   In fig.~\ref{fig:LEEM-HIM} 
   low magnification images of the \GH{Ag/Pt(111)} sample surface recorded
   by LEEM and HIM are shown. The clean Pt(111) is occasionally decorated by
   mounds which sometimes originate from screw dislocations. \GH{These
      mounds are present on the clean Pt(111) surface and are neither formed
      nor affected by the subsequent Ag deposition.} Representative examples
   of such surface mounds are presented in fig.~\ref{fig:LEEM-HIM}. These
   mounds occur after annealing and a relatively swift cooling due to
   surface--bulk mass exchange~\cite{Poelsema2012}. The dark lines (a few
   marked by yellow arrows) in fig.~\ref{fig:LEEM-HIM}(A)
   and~\ref{fig:LEEM-HIM}(B) are steps separating atomically flat terraces.
   Mounds and step bunches are typically accompanied by wide terraces like
   in the top half of the image in fig.~\ref{fig:LEEM-HIM}(A): Step
   migration during the annealing stage is suppressed by the mounds giving
   rise to step bunching and, at the same time, to wider terraces away from
   the obstruction caused by the mound. The mounds are threefold symmetric
   as becomes evident by looking at the relative lengths of the edge
   segments: opposite sides have different lengths, in other words mirror
   symmetry is absent. Moreover, adjacent facets of the mound have different
   brightnesses as is seen clearly in figs.~\ref{fig:LEEM-HIM}(A)
   and~\ref{fig:LEEM-HIM}(B). Unfortunately, field
   distortions~\cite{Nepijko2001} prevent a full structural characterization
   of the mounds by LEEM. Still we can extract important information. Steps
   along $\langle\overline{1}10\rangle$ azimuthal directions on fcc (111)
   surfaces are different. They either have (111) or (001) type microfacets
   of which the former are energetically favoured~\cite{Michely1991}. The
   longer mound edges on opposing sides are apparently energetically
   favoured and from that fact we can derive that the brighter sides of the
   mounds are of the $\left[\overline{1}10\right]$--(111) (oriented
   microfacets) type, while the darker sides in between are of the
   $\left[\overline{1}10\right]$--(100) type. \GH{The threefold symmetry}
   allows us to determine fully the orientation of the mounds in a field
   free situation such as in HIM. The curved dark line on top of t he mound
   represents a monatomic step (marked by yellow arrows in
   fig.~\ref{fig:LEEM-HIM}(B)) as we conclude from the clear correspondence
   with the data in ref.~\citenum{Poelsema2012} and references therein.

\begin{figure}
      \includegraphics{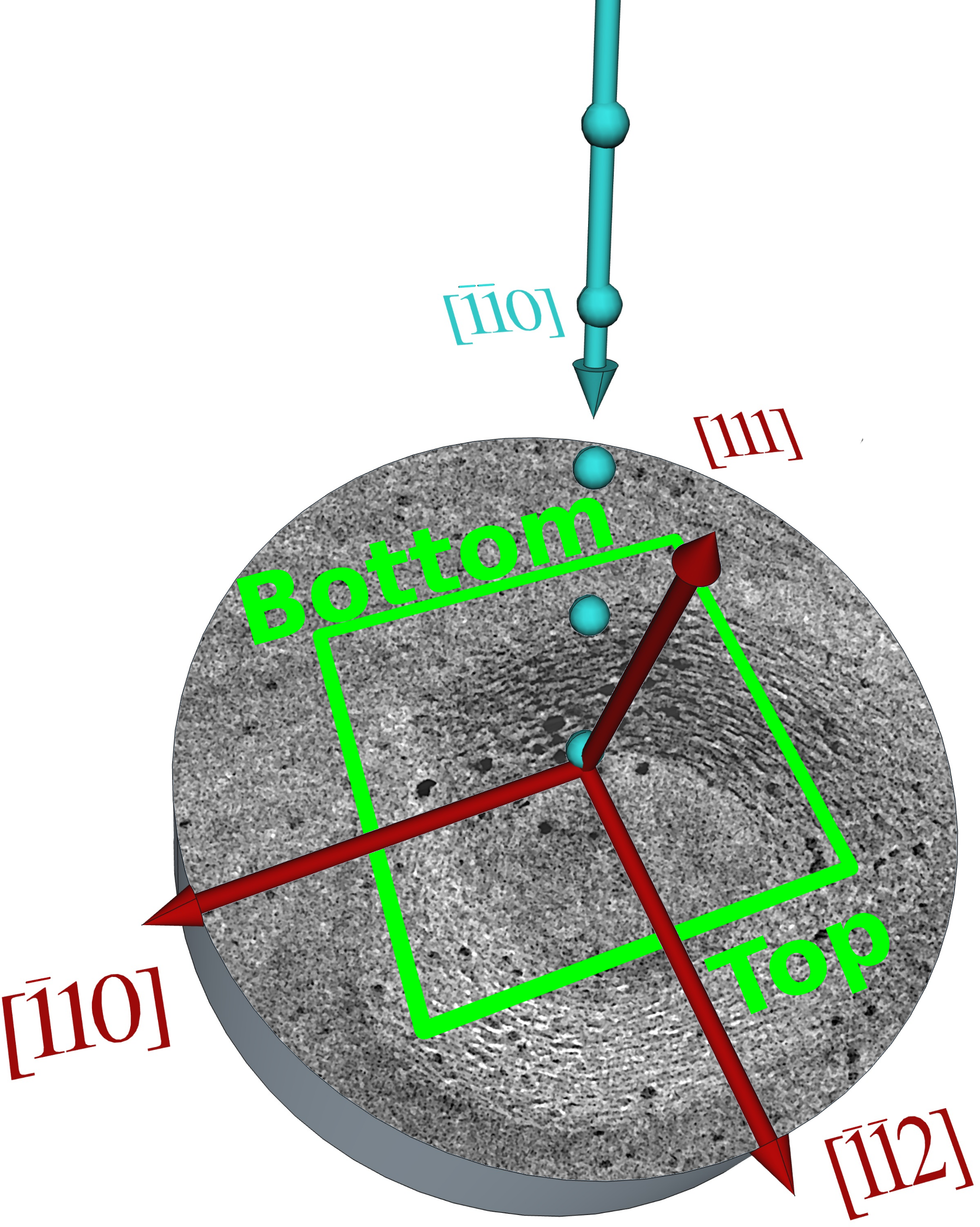}
      \caption{\label{fig:geo} Sketch outlining the geometrical relation
	 between primary He beam (light blue), sample and sample surface
	 directions (red) relevant for the presented work. Please note that
	 the orientation of the recorded images is upside down as indicated by
	 the green labels.}
   \end{figure}

   The appearance of such mounds in HIM can be seen from
   figs.~\ref{fig:LEEM-HIM}(C) and~\ref{fig:LEEM-HIM}(D). A low
   magnification image of a mound is presented in
   fig.~\ref{fig:LEEM-HIM}(C). Note that there is no defined correspondence
   of the azimuth directions in the LEEM and the HIM data. A higher
   magnification of the mound sidewall is shown in
   fig.~\ref{fig:LEEM-HIM}(D). Using HIM images as the one presented in
   fig.~\ref{fig:LEEM-HIM}(C) and fig.~\ref{fig:LEEM-HIM}(D) the orientation
   of the sample with respect to the beam can be determined. While in the
   center right part of fig.~\ref{fig:LEEM-HIM}(D) straight step bunches are
   visible (terminated by $\left\{001\right\}$ facets), irregular and curved
   step bunches are visible at the top and bottom of the image. The straight
   step bunches are preferential oriented parallel to the
   $\langle\overline{1}10\rangle$ directions of the Pt(111) sample surface.
   Using these step bunches the sample has been aligned with the He$^+$ beam
   parallel to the $\left[\overline{11}0\right]$ direction and parallel to
   the Pt$\left(001\right)$ plane. The direction of the arrows in
   fig.~\ref{fig:LEEM-HIM}(C) corresponds to the surface projection
   ($\left\langle\overline{11}2\right\rangle$) of the surface normals of the
   \GH{less abundant} $\left\{001\right\}$ facets. Obviously the
   \GH{dominant} \{111\} oriented step facets are situated just in between.
   In this channelling condition the interaction of the primary beam with
   the sample is minimized and results in a low SE
   yield~\cite{Veligura2012}. As a result the $\left[\overline{1}10\right]$
   direction---except for a small alignment error---is pointing from left to
   right in the subsequent HIM images. As we will see further below, the
   achieved alignment is as good as perfect. The geometrical relationship
   between the primary He beam, the tilted sample and the crystallographic
   directions on the surface is outlined in fig.~\ref{fig:geo}.
   
   \subsection{HIM analysis} 

\begin{figure}
     \includegraphics[width=\linewidth]{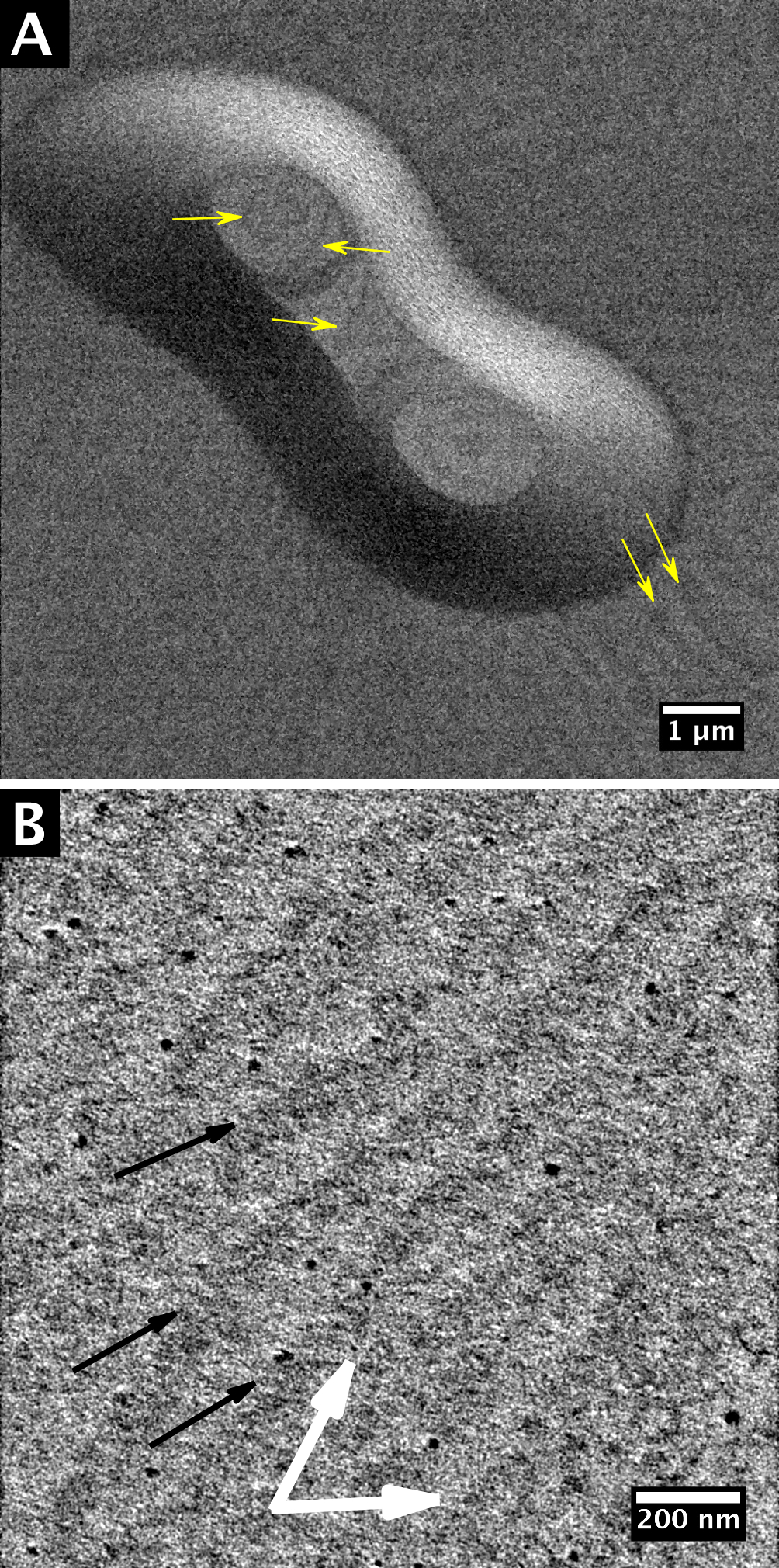}
      \caption{\label{fig:steps}(A) Low magnification HIM image of a mound
	 showing step contrast. The dark curved line on top of the mound
	 present atomic steps. (B) High resolution detail of the Ag/Pt(111)
	 surface showing alternating dark and bright bands. Each pair
	 corresponds to a single terrace. A few step edges are indicated by
	 a black arrows to guide the eye. The horizontal
	 $\left[\overline{1}10\right]$ and the $\left[0\overline{1}1\right]$
	 directions are indicated by white arrows.}
   \end{figure}

   In the low magnification HIM image presented in fig.~\ref{fig:steps}(A) 
   two mounds that originate from a twinned screw dislocation can be seen.
   The dislocation loop intersects the surface at the centers of the mounds.
   On the top of the mound dark bands can be seen. These dark bands have the
   typical appearance of growth steps usually also found at the top of such
   mounds. In the valley the lines meet and annihilate (see top of
   triangles) this can only be explained by a set of coupled screw
   dislocations with opposite rotation sense. The same contrast variation
   can also be observed at the surface surrounding the mounds (see for
   example the right side of fig.~\ref{fig:steps}(A) and
   fig.~\ref{fig:LEEM-HIM}(C)) where micrometer sized terraces are often
   observed (compare also fig.~\ref{fig:LEEM-HIM}(A) for a LEEM image of a
   stepped surface). In fig.~\ref{fig:steps}(B) a high resolution HIM image
   is presented obtained from the flat surface of the Ag/Pt(111) area
   surrounding the mound. An alternating arrangement of dark and bright
   bands can be observed. The $\left[\overline{1}10\right]$ \GH{and
   $\left[0\overline{1}1\right]$ directions are}
   indicated by a white arrows. Ideally steps run along
   $\langle\overline{1}10\rangle$ directions. However, due to local
   curvature variations of the macroscopic surface or the presence of, e.g.,
   mounds they may show quite severe deviations (see
   figs.~\ref{fig:LEEM-HIM}(A) and \ref{fig:LEEM-HIM}(B)). Here the steps
   run diagonally, close to the $\left[0\overline{1}1\right]$ direction but
   not exactly along it. 
  
   The second layer of the alloy consists, according to Secondary Ion Mass
   Spectroscopy~\cite{Ait-Mansour2012} measurements, of 22\% Pt atoms in a
   silver matrix. The constituent of the third layer is mainly Ag.
   Characteristically the third layer propagates by step edges, forming a
   periodically undulated sawtooth--shaped growth
   front~\cite{Jankowski2014b}. Although the shape of the growth front is
   not clearly discernible in fig.~\ref{fig:steps}(B) the black arrows
   indicate the possible location of the single atom high step edges. This
   is based on the \GH{fact} that the \GH{Ag rich} area starts to grow from
   ascending step edges\GH{~\cite{Roeder1993,Jankowski2014b}}. This example illustrates that the
   observation of atomic steps with HIM is not always straightforward for
   reasons not fully understood yet. In this case the discern of the steps
   is attributed to (small) differences in the work function of the second
   and the third layer\GH{~\cite{Jankowski2014b}}.

\begin{figure}
      \includegraphics[width=\linewidth]{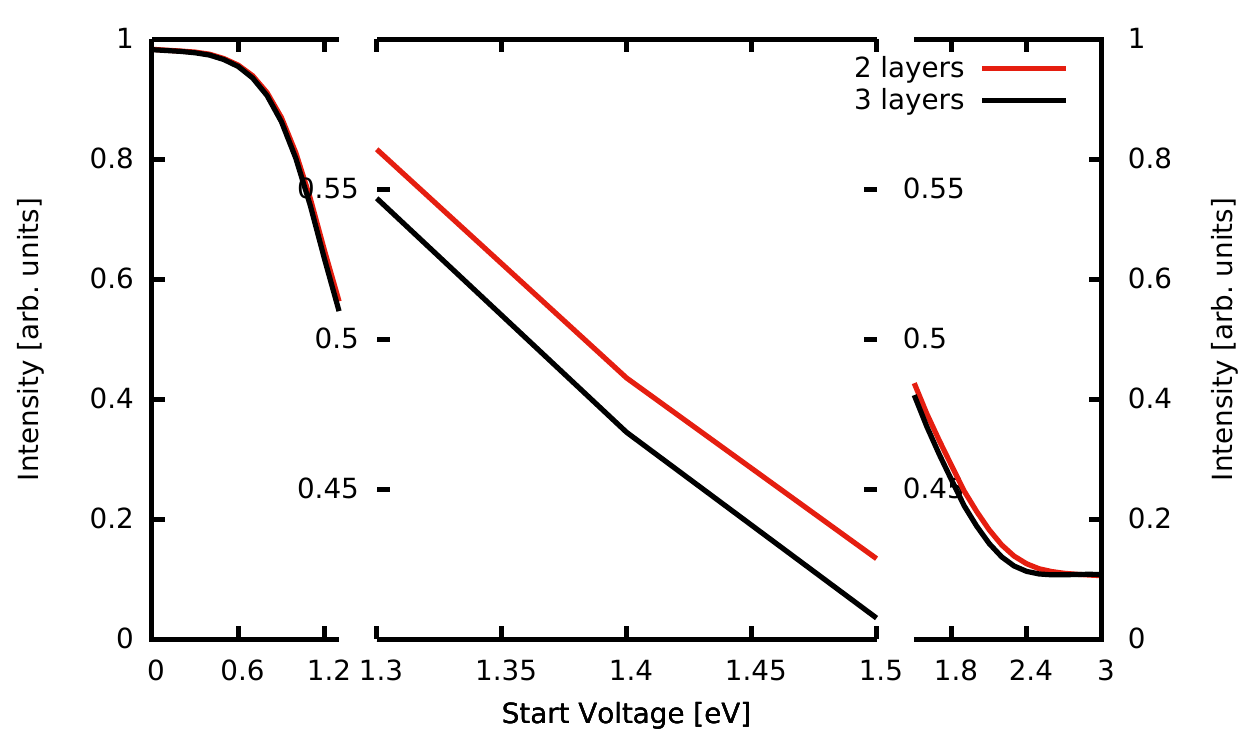}
      \caption{\label{fig:WF} I--V curves recorded from 2 layers of
	 Ag/Pt(111) alloy and 3 layers. The central part of the curve is
	 expanded, both in x and in y--directions. The work function
	 difference (WF) is approximately 25\,meV.} 
   \end{figure}

   The measurement of the relative surface work-function (WF) with LEEM can
   be achieved with high lateral resolution by measuring the local
   brightness variation from a set of images recorded at different start
   voltages. At a very low start voltage, the probing electrons undergo
   total reflection (mirror mode imaging)~\cite{Tromp1993,Bauer1994}
   resulting in uniformly bright images. With increase of the start voltage
   a sharp decrease in the reflected intensity is observed as electrons have
   enough energy to overcome the potential barrier and interact with the
   sample. The value of this critical voltage is a local measure of the
   surface WF. To measure the difference of the surface WF between two
   regions, sets of LEEM images as a function of the start voltage were
   recorded. \GH{The in fig.~\ref{fig:WF} presented bright field intensity
      versus electron voltage (I–V) curve was extracted for each pixel from
      the obtained sets.}
        
   Following the approach discussed above a work function difference between
   the dark and the bright bands of 25\,meV has been obtained (see
   fig.~\ref{fig:WF}). The dark spots visible in figs.~\ref{fig:steps}(B,C)
   are attributed to platinum precipitates found in the Ag/Pt surface alloy
   layer.

\begin{figure*}
      \includegraphics[width=\linewidth]{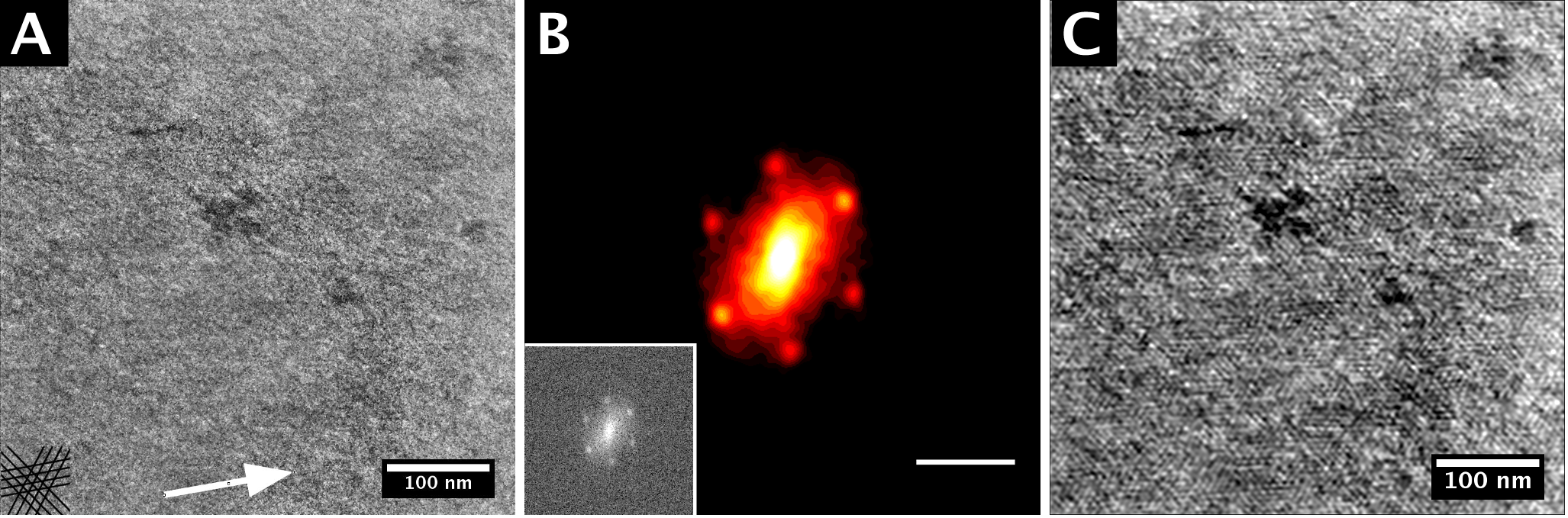}
      \caption{\label{fig:reconstruction}(A) High resolution HIM image of the
	 Ag/Pt(111) surface. Three sets of densely packed parallel lines
	 rotated by 120\textdegree{} can be seen. The dark irregular spots are
	 platinum rich precipitates in the otherwise silver rich surface layer.
	 The white arrow points in the $\left[1\overline{1}0\right]$ direction.
	 (B) FFT of fig.~\ref{fig:reconstruction}(A). The vertical distortion
	 in the FFT is a result of the 35\textdegree{} polar angle between
	 sample surface and image plane. The scale bar corresponds to 5\,nm
	 real space distance measured from the center of the figure. \GH{In the
	    inset the same data is shown using gray scale.} (C)
	 Figure~\ref{fig:reconstruction}(A) filtered with a custom made FFT
	 filter based on fig.~\ref{fig:reconstruction}(B) to enhance the
	 contrast and make the lines more visible.}
   \end{figure*}

   Increasing the magnification further reveals more features of the sample
   surface. \GH{The FOV covers about three or four terraces each of them
      about 100\,nm wide.} Three sets of parallel lines running almost
   horizontally and under $\pm$120\textdegree{} can be seen in the HIM image
   presented in fig.~\ref{fig:reconstruction}(A) and (C) (A set of lines
   under the observed orientation is placed in the lower left corner to help
   the reader). An FFT of the image is provided in
   fig.~\ref{fig:reconstruction}(B).
   Correcting the vertical distortion---which will be discusses later---a 6
   fold symmetric pattern can be observed. A custom made FFT
   filter\GH{\footnote{\GH{A mask based on the FFT presented in
      fig.~\ref{fig:reconstruction}(B) has been used as a bandpass filter in
      the fourier space leaving only the frequencies of interest. The
      remaining FFT has than been converted back into real space to obtain
      fig.~\ref{fig:reconstruction}(C).}}}  based on
   fig.~\ref{fig:reconstruction}(B) has been used to create the contrast
   enhanced version of fig.~\ref{fig:reconstruction}(A) presented in
   fig.~\ref{fig:reconstruction}(C). The spacing of the spots in the FFT is
   at 6.65\,nm$\pm$0.08\,nm after applying the necessary correction due to
   the tilt of the sample in the vertical direction (see below for further
   discussion). From the orientation of the spot pattern it is clear that
   this distance is along the $\langle\overline{11}2\rangle$ set of
   directions. Please note that both the white arrow in
   fig.~\ref{fig:reconstruction}(A)---indicating the
   $\left[1\overline{1}0\right]$ direction---as well as the FFT pattern in
   fig.~\ref{fig:reconstruction}(B) are slightly rotated off the horizontal
   due to a small azimuthal misalignment of the sample (please see
   fig.~\ref{fig:geo} for a sketch of the used imaging geometry). This
   azimuthal misalignment amounts to only 3.6\textdegree{}. Not only the
   misalignment of the azimuthal angle but also the possible misalignment of
   the polar angle of incidence of the ion beam can be checked
   quantitatively. The vertical distortion is a direct result of the
   off-normal direction of the beam (polar angle $\Theta$). As a consequence
   the ion beam travels a longer distance upon applying a deflection voltage
   in the plane of incidence than upon applying the same voltage
   perpendicular to the plane of incidence by a factor of $1/\cos\Theta$.
   This makes the in--plane dimensions of the structures seem shorter by a
   factor of $\cos\Theta$. Obviously, in reciprocal space the pattern gets
   expanded in the vertical direction as is the case in
   fig.~\ref{fig:reconstruction}(B). Since the off--normal incidence of the
   ion beam is the only factor that could destroy the threefold symmetry of the image one can restore the symmetry by
   shrinking the vertical axis by $\cos\Theta$. An optimal symmetry is
   obtained for $\Theta$=36.9\textdegree{}--38.6\textdegree{}, i.e. only
   1.6\textdegree{}--3.3\textdegree{} off the ideal angle. We conclude that
   the realized alignment is excellent. However, due to the relatively low
   beam energy of 30\,keV the acceptance angle for channeling is
   sufficiently large to allow for an effective channeling through the first
   few nanometers~\cite{Veligura2012}.

   \subsection{AFM data}

\begin{figure} 
      \centering \includegraphics[width=\linewidth]{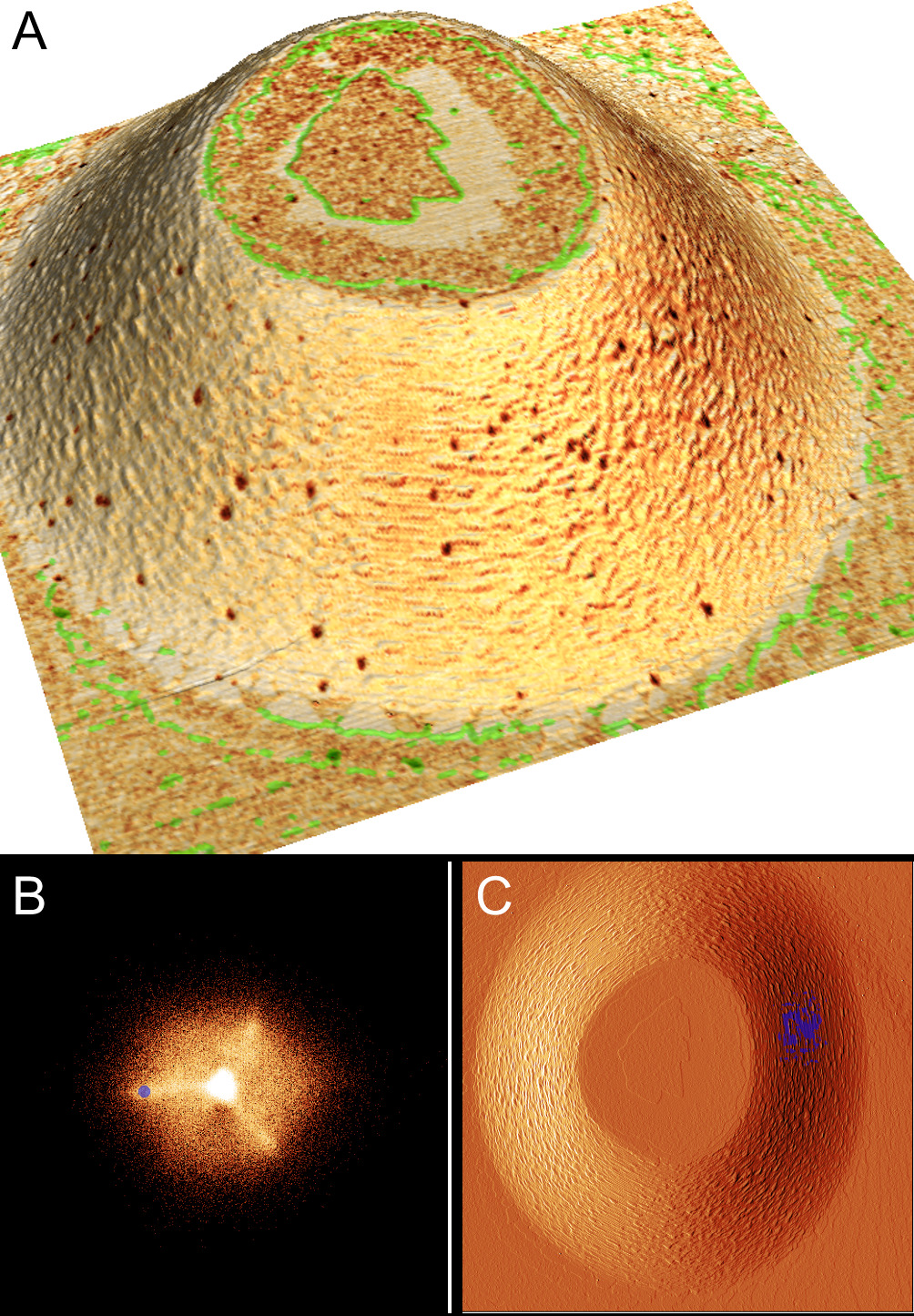}
      \caption{\label{fig:afm} (a) 3D projection of a 120\,nm high surface
	 mound recorder with AFM. The topography data has been shaded using
	 the phase signal to reveal changes in surface termination. Step
	 edges extracted from the amplitude signal are indicated by green
	 lines. (FOV: 5\textmu{}m) \GH{(b) Slope distribution obtained from
	    the topography data. A small selection of slopes is marked in
	    blue. Slopes range from $-0.35$ to $+0.35$ (c) Amplitude image of the mound presented in (a). The
	    blue colored straight steps on the right side of the mound
	    correspond to the slopes selected in (b). (FOV; 5\,\textmu{}m)}}
   \end{figure}

   AFM data of a mound, recorded ex--situ after the deposition of 3 layers of
   Ag, are presented in fig.~\ref{fig:afm}. 
   The 3D representation of the topography data, shown in
   fig.~\ref{fig:afm}(a), is shaded using the phase signal. It confirms the
   HIM and LEEM observations. The mound and its surrounding are decorated by
   small black spots. They are attributed to the platinum--rich precipitates
   also visible in HIM and LEEM images (see fig.~\ref{fig:steps}(B)--(C) for
   comparison) as dark objects. Formation of those precipitates was observed
   in previous investigations~\cite{Vroonhoven2005,Jankowski2014b} upon
   completion of the first layer, and were identified there, as few atoms
   high Pt clusters embedded in the surrounding. The clusters become less
   abundant but persist on the surface until a few tens of \AA{}ngstr\"om of
   Ag are deposited. The key observation from the data presented in
   fig.~\ref{fig:afm}(a) is that the top of the mound shows regions
   corresponding to two different phases. One is the area appearing dark in
   the AFM phase data which occupies the center and rim area of the mound,
   the other phase is represented in white and found in between the dark
   areas. The fact that we can observe Pt clusters embedded in both types of
   regions, means that both regions correspond to at least one layer of the
   alloy. The observed step and terrace pattern on top of the mound is quite
   characteristic for the growth of the first few layers of
   Ag/Pt(111)~\cite{Vroonhoven2005,Jankowski2014b}. The higher coverage on
   top of the mound is explained by steering induced flux enhancement at
   protrusions~\cite{Dijken1999}. The diffuse border separating the bright
   (Pt--rich) area from the outer dark (Ag--rich) area represents the growth
   front of the surface alloy. \GH{In figure~\ref{fig:afm}(b) the slope
      distribution is presented. The slope distribution graph plots the
      local derivative of the surface and creates a 2D histogram from
      it~\cite{Klapetek}. The hexagonal shape of the mound is visible if one
      looks at the overall shape of the distribution. The three spokes that
      can be seen at 1, 5 and 9 o'clock represent areas with equal
      orientation in a very narrow azimuth angle band ($<$1\textdegree{}).
      These are the regular step bunches that are straight and not curved.
      They can also be seen in the amplitude image (fig.~\ref{fig:afm}(c)).
      There small bands of straight steps can be seen at 3, 7, and 11
      o'clock. These straight step segments are responsible for the spokes
      in the slope distribution. In these areas the polar angle changes
      slightly (at the top and bottom of the mound, giving rise to the spoke
      in fig.~\ref{fig:afm}(b)) but the azimuth angle is constant. As an
      example the facets at the 3 o'clock side of the mound are highlighted
      in both figures ($\approx$7\textdegree{} polar,
      $\approx$3.7\textdegree{} azimuth $\pm$0.25\textdegree). This is one
      of the $\left\langle\overline{11}2\right\rangle$ directions. The
      straight steps run parallel to one of the
      $\left\langle\overline{1}10\right\rangle$ directions.} 

   \section{Discussion and conclusions}

   Helium Ion Microscopy is well known for its surface
   sensitivity~\cite{Hlawacek2013b}. The usually obtained SE images are
   sensitive to small variations in work
   function~\cite{George2012,Hlawacek2012,Buchholt2011}. Both effects are
   related to the particularities of the SE generation in
   HIM~\cite{Ohya2009,Petrov2011}. The work function difference (see
   fig.~\ref{fig:WF}) between the
   platinum rich (dark) and silver rich (bright) areas (see
   fig.~\ref{fig:steps}(A)--(B)) is only 25\,meV. This is
   a remarkable example for both the surface sensitivity as well as the
   susceptibility of the method to changes in the work function.
   Although the steps can not be seen
   directly from the morphology this is the first visualization of a single
   atomic layer
   high steps in a SE HIM image. Previous---topography based---images of
   steps~\cite{Buchholt2011} are obtained from materials with large unit cells
   where each step is several atoms high.

\begin{figure}
      \includegraphics[width=\linewidth]{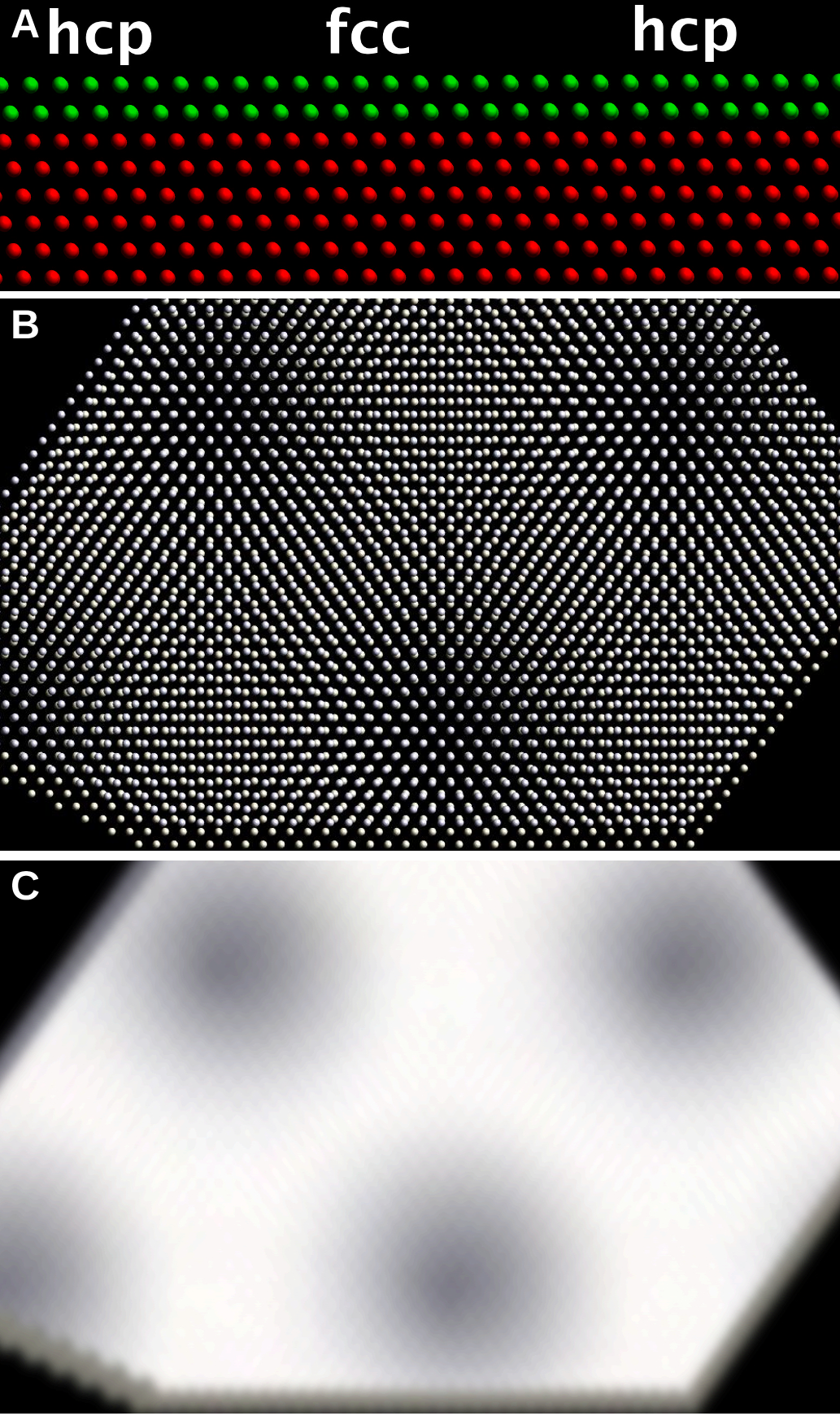}
      \caption{\label{fig:model} Atomistic model of 2 layers Ag on Pt(111). (A)
	 side view along $\left[\overline{1}10\right]$. In this projection
	 23 Ag atoms (green) cover 24 Pt atoms (red). Fcc
	 and hcp areas are indicated. (B) View on the crystalline surface
	 along the beam. (For clarity all atoms are colored grey) (C) Expected qualitative SE-yield distribution.}
   \end{figure}

   The results presented in fig.~\ref{fig:reconstruction} are a
   demonstration of the applicability of the dechanneling
   contrast~\cite{Hlawacek2012}. The 2 layers Ag on Pt(111) system is
   characterized by a stress induced triangular network of dislocations originating from the interface.
   These dislocations separate areas in the first two layers which are
   shifted along the [$11\overline{2}$] direction over $1/\sqrt{3}$ of the
   nearest neighbour distance. Locally, this leads to a different
   stacking of the layers as compared to the bulk fcc stacking. In
   fig.~\ref{fig:model}(A) a side view of the simplified atomic structure is
   presented. This is not the actual atomic structure of Ag/Pt system but
   should clarify why we can obtain contrast on such a surface. For a more
   accurate description of the atomic model the reader is referred
   to ref.~\citenum{Ait-Mansour2012}.
   In this projection the first two layers are comprised of 23 Ag atoms
   while the underlying bulk Pt contains 24 atoms over the same distance. As
   a consequence a fcc stacking is observed in the center of the image while
   a hcp stacking is observed at the left and right side. In the latter
   areas the silver atoms suppress axial channeling of the helium ions into
   the Pt. A detailed characterization of this system with respect to the
   surface can be found in~\cite{Jankowski2014, Ait-Mansour2012}. Using the
   step bunches on the mounds (see fig.~\ref{fig:LEEM-HIM}(C) and (D)) the
   sample has been aligned azimuthally with the
   $\left[\overline{1}10\right]$ direction roughly parallel to the tilt axis
   direction. Using a tilt of 35\textdegree{} the substrate has been
   oriented in a $\langle\overline{11}0\rangle$ channeling direction for the
   incoming He$^+$ beam. In hcp areas the top two atomic layers of the
   surface are shifted by two thirds of the $\langle\overline{1}10\rangle$
   inter chain spacing leading to local CAC (hcp) stacking in stead of local
   ABC (fcc) stacking. This closes the channel and results in an increased
   scattering probability. In fig.~\ref{fig:model}(B) the view of the
      beam on the atomic structure is presented. The 6--fold symmetric
      arrangement of the fcc channeling areas is visible. Subsequently, a
   higher SE yield is obtained in such hcp areas compared to the fcc areas
   with a lower scattering probability. A qualitative representation of
      the expected SE yield is presented in fig.~\ref{fig:model}(C). The
   value of 6.65\,nm$\pm$0.08\,nm for the periodicity of the surface pattern
   is in excellent agreement with the 6.55\,nm measured by
   SPA--LEED~\cite{Jankowski2014}. Both values correspond to the distance
   between parallel base vectors of the super cell. The numbers show that
   the calibration of the HIM is correct to about 1\%. The obtained
   periodicity of 6.65\,nm agrees with a situation in which 23 Ag--atoms in
   the top--layer reside on 24 Pt lattice sites to an accuracy of 0.03\%.
   Rightfully we can and do claim that we reach atomic precision here. We do
   not claim atomic resolution, since that would require the imaging of
   individual atoms. We like to suggest, however, that it is possibly not
   the instrument’s limitation but rather insufficient atomic contrast which
   prevents atomic resolution.

   In summary we have---to the best of our knowledge---for the first time
   visualized single atom high surface steps using Helium Ion Microscopy. The
   contrast is based on minute changes of the work function across the
   otherwise atomically flat terraces. In addition we used the dechanneling
   contrast mechanism to successfully visualize the periodic arrangement of fcc
   and hcp regions in a 2--3 layers thick surface film. This pattern is the result of
   a triangular dislocation network at the interface between the silver
   alloy 
   layer and the platinum bulk. The contrast in HIM is the result of 
   a lateral displacement of the outermost film layers over a nearest
   neighbour distance divided by $\sqrt3$ along a
   $\langle11\overline{2}\rangle$ direction.
   The measured
   periodicity of 6.65\,nm and the fact that the contrast is based on
   channeling/dechanneling confirms in detail earlier measurements using other methods.
   The obtained results are powerful demonstrations of the surface sensitivity and 
   high resolution capabilities of the Helium Ion Microscopy. 

   \section*{Acknowledgements}

   We want to thank Robin Berkelaar for acquiring the AFM data. This work is
   part of ECHO research program 700.58.026, which is financed by the
   Chemical Sciences Division of the Netherlands Organisation for Scientific
   Research (NWO).

   \bibliographystyle{elsarticle-num}
   \bibliography{him_version009}

\end{document}